\documentclass[sigplan,10pt]{acmart}
\setcopyright{none} 
\renewcommand\footnotetextcopyrightpermission[1]{} 

\AtBeginDocument{%
  }

\begin{document}

\title{The Kernel's Write: Application Read-Only Memory}
\pagestyle{plain} 

\author{Hui Sub Shim}
\affiliation{%
  \institution{Stanford University}
  \city{Stanford}
  \state{California}
  \country{USA}}
\email{hsshim@stanford.edu}

\author{Katherine Mohr}
\affiliation{%
  \institution{Stanford University}
  \city{Stanford}
  \state{California}
  \country{USA}}
\email{kmohr@cs.stanford.edu}

\author{Philip Levis}
\affiliation{%
  \institution{Stanford University}
  \city{Stanford}
  \state{California}
  \country{USA}}
\email{pal@cs.stanford.edu}

\renewcommand{\shortauthors}{Shim et al.}

\begin{abstract}
Alongside power, DRAM has become a major limiting factor in datacenter growth.
As DRAM's cost-per-bit has plateaued over the past decade, a class of emerging memory technologies, 
called Long-term RAM (LtRAM), offers a path to denser and cheaper main memory. 

However, LtRAM has three main drawbacks: asymmetric read\slash write latencies, limited
endurance, and coarse write granularity. 
In an attempt to isolate software from these drawbacks, LtRAM technologies such as Intel Optane 
copy an approach from flash devices and introduce a translation layer
that manages wear-leveling, address remapping, and read\slash write caching.
Prior experimental studies have found these operations add significantly to LtRAM latency.

Rather than making LtRAM look like DRAM, we propose redesigning the hardware\slash software interface to offload more responsibility to the operating system.
This design hinges on one central property, Application Read-Only Memory (AROM): LtRAM pages are read-only to applications and written only by the OS during page migrations.
AROM is enforced by leveraging copy-on-write (CoW): application writes to LtRAM trigger a fault that migrates the page back to DRAM before the store is applied.
This invariant allows us to shift LtRAM management from the on-DIMM controller to the operating system, drastically simplifying the DIMM's hardware.
With this approach, we aim to match the performance of pure DRAM on read-mostly workloads while delivering LtRAM's density and cost advantages.


\end{abstract}

\maketitle

\section{Introduction}
\label{sec:intro}



Memory has become the dominant share of server cost.
Microsoft Azure and Meta both report that DRAM accounts for over half of total server cost~\cite{pond2023,tpp2023}, a problem that has accumulated as DRAM cost-per-bit has plateaued~\cite{li_towards_2025}.

Long-term RAM (LtRAM) is a recent proposal to mitigate this issue.
LtRAM is a class of emerging memory technologies that compete with DRAM's read latency, but with lower per-bit cost.
Some examples include RRAM, PCM, FeFET, and MRAM.
Prior work has identified LtRAM as well-suited for long-lived, read-mostly data~\cite{mrm2025}.
Replacing the DRAM that holds such data with LtRAM has the potential to reduce DRAM footprint and overall main memory cost.

However, LtRAM brings its own host of challenges: read\slash write latencies are asymmetric, endurance is limited, and write granularity is coarse.
Intel Optane, a discontinued phase-change memory (PCM) module, is the canonical real-world example of LtRAM. 
To minimize software-level changes, Optane adopted the translation-layer pattern from flash solid state drives (SSD), pushing wear-leveling, address remapping, and read\slash write caching into on-DIMM hardware.
This hardware complexity added latency on top of the raw media access time and inflated module cost, diminishing the performance and cost advantages of the underlying technology.

We argue that forcing DRAM compatibility on LtRAM adds too much hardware complexity, eroding the performance and cost advantages LtRAM can deliver.
A new interface between the OS and memory would let the OS manage LtRAM properties with metadata it already has, rather than hiding them in the module.
With this new interface, we propose a mixed DRAM and LtRAM main memory that aims to match pure-DRAM performance while reducing per-server DRAM footprint and overall main memory cost.
Our latency projection places LtRAM 26--79\% faster than Optane and between 10\% faster and 220\% slower than DRAM on per-access read latency, while providing density and cost benefits.

This paper makes four contributions:
\begin{enumerate}
  \item An analysis of the performance hits incurred by Intel Optane's on-module translation layer (\S\ref{sec:optane}).
  \item The \textbf{Application Read-Only Memory (AROM)} invariant: only the OS writes to LtRAM, enforced by copy-on-write (CoW) on application stores (\S\ref{sec:interface}).
  \item A novel interface that uses AROM to move LtRAM management from the on-DIMM controller to the OS, avoiding Optane's controller overheads (\S\ref{sec:interface}).
  \item A prototype design on the Enzian~\cite{enzian} platform with NOR flash as the LtRAM (\S\ref{sec:implementation}).
\end{enumerate}
\section{Motivation: The Problem with DRAM}
\label{sec:motivation}

DRAM's share of server cost has grown for two reasons. Memory demand has increased with core counts and per-server working-set sizes, while DRAM cost-per-bit has plateaued for over a decade. DRAM cell density may continue to improve through emerging technologies such as vertical-channel access transistors~\cite{ha2023vct} and 3D-stacked DRAM~\cite{han2023stackeddram}. However, manufacturing constraints on capacitor scaling prevent cost-per-bit from dropping~\cite{li_towards_2025}. The 2025--2026 AI demand surge has exacerbated this trend, with prices for mainstream DIMMs roughly doubling in 1Q26 alone~\cite{trendforce2026q1,counterpoint2026}. 

The DRAM price spike from the AI demand surge may resolve as suppliers expand capacity, but the underlying cost-per-bit plateau will not. The plateau reflects fundamental technological limits, not market dynamics. DRAM's share of server cost is therefore projected to keep growing.

To address this unsustainable growth, software must use DRAM more efficiently, or hardware must adopt cheaper memory alternatives. Software techniques such as memory compression~\cite{sdfm2019} and page deduplication slow DRAM usage growth but cannot lower the cost-per-bit. Memory tiering~\cite{tpp2023,colloid2024,alto2025} demotes cold pages to slower, cheaper hardware alternatives such as Compute Express Link (CXL)~\cite{zhong2024cxl}, reusing previous-generation DRAM or new memory devices. The next section explores an alternative approach: LtRAM. 

\section{Long-term RAM (LtRAM)}                 
\label{sec:ltram}
Long-term RAM (LtRAM) is a recently proposed class of memory technologies intended as a denser, cheaper complement to DRAM~\cite{mrm2025}. LtRAM matches DRAM's read latency and reaches densities several times those of current DRAM, with associated reductions in cost-per-bit. However, this density comes with tradeoffs: writes are slower, write granularity is coarser than a cache line, and endurance is bounded. 

Given these write restrictions, LtRAM is best suited for long-lived, read-mostly data: code pages, machine learning inference weights~\cite{mrm2025}, and in-memory stores such as Redis and Memcached~\cite{li_towards_2025}. Substituting LtRAM for the DRAM that holds such data can lower per-server memory cost.

To make this concrete, we survey the candidate LtRAM technologies and the design space they expose (\S\ref{sec:ltram-space}) and analyze how Intel Optane's controller mechanisms sacrificed those gains in practice (\S\ref{sec:optane}).
\subsection{The LtRAM design space}
\label{sec:ltram-space}

\begin{table}[!t]
  \small
  \setlength{\tabcolsep}{4pt}
  \begin{tabular}{lrrcll}
    \hline
    Device & Read & Write & Endur- & Write & Cost \\
           & lat. & lat.  & ance   & gran. & per bit \\
    \hline
    DRAM          & 80\,ns  & 80\,ns       & $\infty$  & cache line & high \\
    3D V-RRAM     & 100\,ns & 1\,$\mu$s    & $10^6$    & page       & low \\
    3D FeFET      & 200\,ns & 10\,$\mu$s   & $10^5$    & page       & low \\
    PCM           & 300\,ns & 1\,$\mu$s    & $10^8$    & 256\,B     & mid \\
    STT-MRAM      & 20\,ns  & 20\,ns       & $10^{15}$ & word       & high \\
    \hline
  \end{tabular}
  \caption{Representative LtRAM memories and their characteristic shape, with DRAM as reference. Numbers should be read for shape rather than precision, as each row can shift by orders of magnitude with material and circuit choices~\cite{lu_high_2024}}
  \label{tab:ltram}
\end{table}

Table~\ref{tab:ltram} summarizes four LtRAM technology candidates~\cite{lu_high_2024}.
Although the candidates differ in physical mechanism, they generally share the same shape: large read\slash write latency asymmetry, coarse write granularity, 
and hard endurance limits. LtRAM is the abstraction over this shape.
These memory technologies can continue density and cost-per-bit scaling where DRAM has stalled~\cite{mrm2025}. 


\begin{table*}[!t]
  \footnotesize
  \setlength{\tabcolsep}{4pt}
  \begin{tabular}{p{0.22\textwidth}p{0.35\textwidth}p{0.35\textwidth}}
    \hline
    \textbf{Optane Mechanism} & \textbf{Resulting Overhead} & \textbf{LtRAM Interface Fix} \\
    \hline
    Cache-line Granularity Mismatch (\S\ref{sec:cgm}) & 75\% write throughput drop, 2$\times$ random-read latency & Cache-line reads, page writes; \newline OS page management naturally coalesces writes \\[3pt]
    Wear-Leveling (\S\ref{sec:wearleveling}) & Opaque migration fires every $\sim$3.4\,MB written; \newline $\sim$60\,$\mu$s tail read latency per event ($>$160$\times$ latency) & OS-side wear-leveling; \newline hardware has no migration path \\[3pt]
    Address Indirection Table (\S\ref{sec:ait}) & 76--200\,ns added to every read & OS-side wear-leveling; no on-module map \\[3pt]
    Workload-Agnostic Design (\S\ref{sec:target-workload}) & 70\% read bandwidth drop under 50/50 R/W; \newline 4$\times$ more sensitive to mixed workload than DRAM & Read-mostly by design; \newline OS-paced writes via token allocator \\
    \hline
  \end{tabular}
  \caption{Each Optane mechanism carries an overhead that the LtRAM interface eliminates by shifting responsibility to the OS.}
  \label{tab:optane}
\end{table*}

\subsection{Case Study: Intel Optane}
\label{sec:optane}
Intel Optane, the only LtRAM-class memory ever available as a DIMM, illustrates what goes wrong when non-DRAM memory is hidden behind a DRAM interface.

Each of Optane's controller-side mechanisms was introduced to hide memory complexity from software. Hiding that complexity comes at a measurable cost: a read-modify-write on every sub-256$\,$B store, an AIT lookup on every read, a 160$\times$ tail-latency spike from internal wear-leveling migrations, and a bandwidth collapse under mixed read\slash write workloads. 
Table~\ref{tab:optane} summarizes four of Optane's hardware mechanisms and their resulting overheads.
After a brief overview of Optane, the following subsections address each of its hardware overheads in turn.

\subsubsection{Intel Optane Background}

Intel Optane DC Persistent Memory Module (DCPMM) is a byte-addressable non-volatile DIMM that sits on the memory bus alongside DRAM and exposes 3D-XPoint media to the CPU at cache-line granularity.
The on-DIMM controller (XPController) mediates between cache-line granularity transactions from the CPU and the coarser-granularity 3D-XPoint media, shown in Figure~\ref{fig:optane-internals}.
Three structures drive Optane's access latency:
\begin{figure}[t!]
  \centering
  \includegraphics[width=0.85\columnwidth]{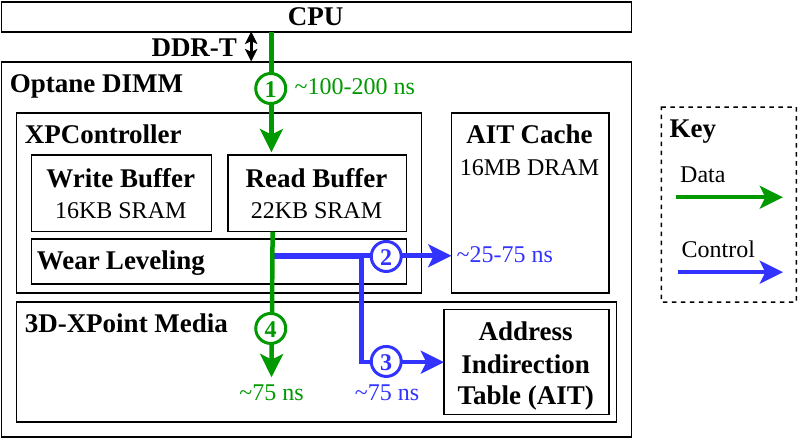}
  \caption{Optane internals, with a sample read path. Worst-case reads traverse the read buffer, AIT cache, and media-resident AIT before fetching data from the 3D-XPoint media. Numbers are approximate based on prior work~\cite{wang_micro20, liu2023side-channel}.}
  \label{fig:optane-internals}
\end{figure}

\textbf{3D-XPoint media.} 3D-XPoint is the phase-change memory (PCM) media that backed Optane. It has a 256\,B access granularity, which Intel calls an XPLine.

\textbf{Address Indirection Table (AIT) and cache.} The controller maintains a host-invisible map from CPU physical addresses to media locations. This map implements hardware wear-leveling and bad-block remapping. The full AIT lives in the media, and a 16\,MB on-DIMM DRAM buffer caches recent entries at 4\,KB granularity~\cite{izraelevitz2019basic, wang_micro20, optane2022}.


\textbf{Write and Read buffers.} The DIMM controller maintains a small write-combining buffer ($\sim$12--16\,KB) that coalesces sub-XPLine writes and a read buffer ($\sim$16--22\,KB) that caches XPLines to speed up sequential reads~\cite{wang_micro20, optane2022}.

These components introduce additional complexity to Optane's hardware, increasing manufacturing costs and making Optane's performance highly sensitive to workload.

\subsubsection{Cache-line Granularity Mismatch}
\label{sec:cgm}

While modern processors have 64\,B cache lines, the 3D-XPoint memory technology naturally handles reads and writes in XPLines.
Consequently, random cache-line writes suffer, as sub-XPLine writes must perform a read-modify-write (RMW) operation, reading the XPLine from media, modifying it, and writing it back.
While sequential writes can be coalesced in a write-combining buffer, random 64\,B writes incur a 4$\times$ write amplification, dropping throughput from $\sim$2.3\,GB/s to $\sim$0.56\,GB/s ($\sim$75\% degradation)~\cite{wang_micro20}.

Random cache-line reads face the same penalty.
The controller hides read latency of sequential 64\,B reads by fetching whole XPLines into an on-DIMM read buffer, achieving $\sim$169\,ns sequential read latency. 
Meanwhile, random reads incur 4$\times$ amplification and $\sim$305--374\,ns latency~\cite{yang2020empirical,hirofuchi2020prompt}.

\subsubsection{Wear-Leveling}
\label{sec:wearleveling}

Optane's controller performs wear-leveling internally. When writes concentrate on a small region, the controller migrates data off the worn cells.
Wang et al.~\cite{wang_micro20} measure that XPLine overwrites on Optane trigger such a migration roughly every 14{,}000 writes (about 3.4\,MB). Tail latencies reach roughly 60\,$\mu$s per read request, over 160$\times$ the typical random-read latency.

\subsubsection{Address Indirection Table}
\label{sec:ait}

Intel Optane uses an Address Indirection Table (AIT) and corresponding cache for wear-leveling and bad-block management.
Every read consults the AIT before reaching the media. Liu et al.~\cite{liu2023side-channel} measure reads with AIT-cache hits at 351\,ns and misses at 427\,ns. The 76\,ns gap is the extra media access required on a miss to fetch the AIT entry.
On a 128\,GB DIMM, the 16\,MB AIT cache covers only $\sim$8\,GB of address space, or $\sim$6\% of the device (assuming standard cache metadata: valid, tag, LRU, dirty, ECC). Therefore, 94\% of random accesses miss the AIT cache and pay the 427\,ns worst-case path~\cite{liu2023side-channel, wang_micro20, izraelevitz2019basic, optane2022}.



\subsubsection{Workload-Agnostic Design}
\label{sec:target-workload}

Optane was designed as a memory-bus-attached DIMM, intended to support the same read\slash write workload mix as DRAM. Microbenchmark measurements show that DRAM sustains its bandwidth across all read\slash write mixes, while Optane does not. Izraelevitz et al.~\cite{izraelevitz2019basic} measure that Optane's aggregate sequential bandwidth collapses from $\sim$30\,GB/s on pure reads to $\sim$10\,GB/s under a 50/50 mixed workload, a 67\% loss in read throughput.

\section{Proposed Hardware/Software Interface}
\label{sec:interface}

\begin{figure}[t!]
  \centering
  \includegraphics[width=0.87\columnwidth]{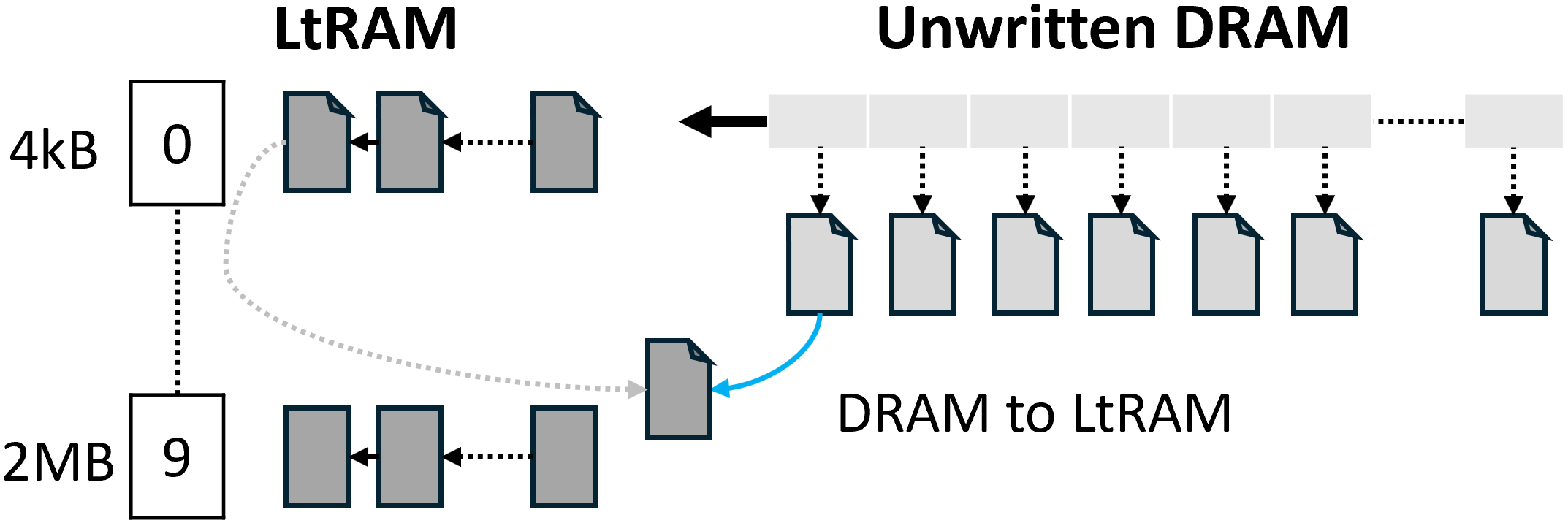}
  \caption{DRAM to LtRAM Migration. LtRAM is a distinct memory zone with its own free list. Pages exceeding a write-recency threshold are migrated from DRAM to LtRAM via Linux's existing page management.}
  \label{fig:ltram-internals}
\end{figure}

Optane relied on complex hardware logic to present a simple, DRAM-compatible interface to software, but each of these mechanisms added measurable performance overheads.

We introduce a novel HW\slash SW interface that enforces an invariant we call \textbf{Application Read-Only Memory (AROM)}. 
Under AROM, LtRAM is read-only to applications, and the kernel may only write to LtRAM when migrating pages from DRAM. This design relocates LtRAM management from the on-DIMM controller to the OS. The device-side hardware exposes reads at cache-line granularity, writes at 4\,KB page granularity, and cell health metadata; the OS manages all policy decisions with its existing structures.

\subsection{Controller Requirements}
\label{sec:interface_controller}

The device-side controller has three responsibilities:
\begin{enumerate}
    \item accept reads at cache-line granularity,
    \item accept writes at 4\,KB page granularity, and
    \item expose each memory block's health to the OS.
\end{enumerate}

Reads remain at cache-line granularity, where LtRAM technologies achieve read latency competitive with DRAM.
The write granularity is set to 4\,KB, a multiple of every LtRAM technology's write granularity (Table~\ref{tab:ltram}).
Writes are therefore aligned to the memory's boundary, structurally eliminating the read-modify-write path that drove Optane's 75\% write-throughput collapse (\S\ref{sec:cgm}).
The controller reports per-block health to the OS, which uses this to seed its wear-leveling state with each block's current wear level.

The controller does not implement wear-leveling. This eliminates the AIT and its overheads:
the AIT cache lookup and the $\sim$75\,ns media access incurred on the 94\% of random reads that miss the AIT cache (\S\ref{sec:ait}). It also removes the $\sim$60\,$\mu$s tail-latency spikes from hardware-driven wear-leveling migrations (\S\ref{sec:wearleveling}). Only the memory's intrinsic access latency remains, surfaced through a simpler controller without an AIT or its AIT DRAM cache.

\subsection{OS Requirements}
\label{sec:interface_os}

The operating system provides four guarantees:
\begin{enumerate}
    \item the only writes issued to LtRAM are 4\,KB page writes from the kernel migration path;
    \item application stores to an LtRAM-resident page trigger a copy-on-write (CoW) fault that migrates the page back to DRAM before the store applies;
    \item pages migrated into LtRAM are read-mostly, avoiding the mixed-workload bandwidth collapse (\S\ref{sec:target-workload}); and
    \item write traffic to LtRAM is balanced across cells, replacing the controller's wear-leveling (\S\ref{sec:wearleveling}).
\end{enumerate}

Guarantees (1) and (2) enforce AROM. LtRAM is read-only from the application's perspective; only the kernel migration thread writes to it, and only at 4\,KB granularity. 
Aligning to 4\,KB lets the OS manage LtRAM through existing page mechanisms.
Application traffic also never reaches the media at sub-page granularity, so the controller never incurs the read-modify-write path. Guarantees (3) and (4) move migration policy and wear leveling from the controller to the OS: page migrations are handled by a kernel policy that classifies pages by write frequency, and endurance is managed by the OS at a page granularity (e.g., via a token allocator). \S\ref{sec:implementation} describes possible implementations of both.

%
%

\section{Implementation}        
\label{sec:implementation}
\begin{figure}[t]
  \centering
  \includegraphics[width=0.8\columnwidth]{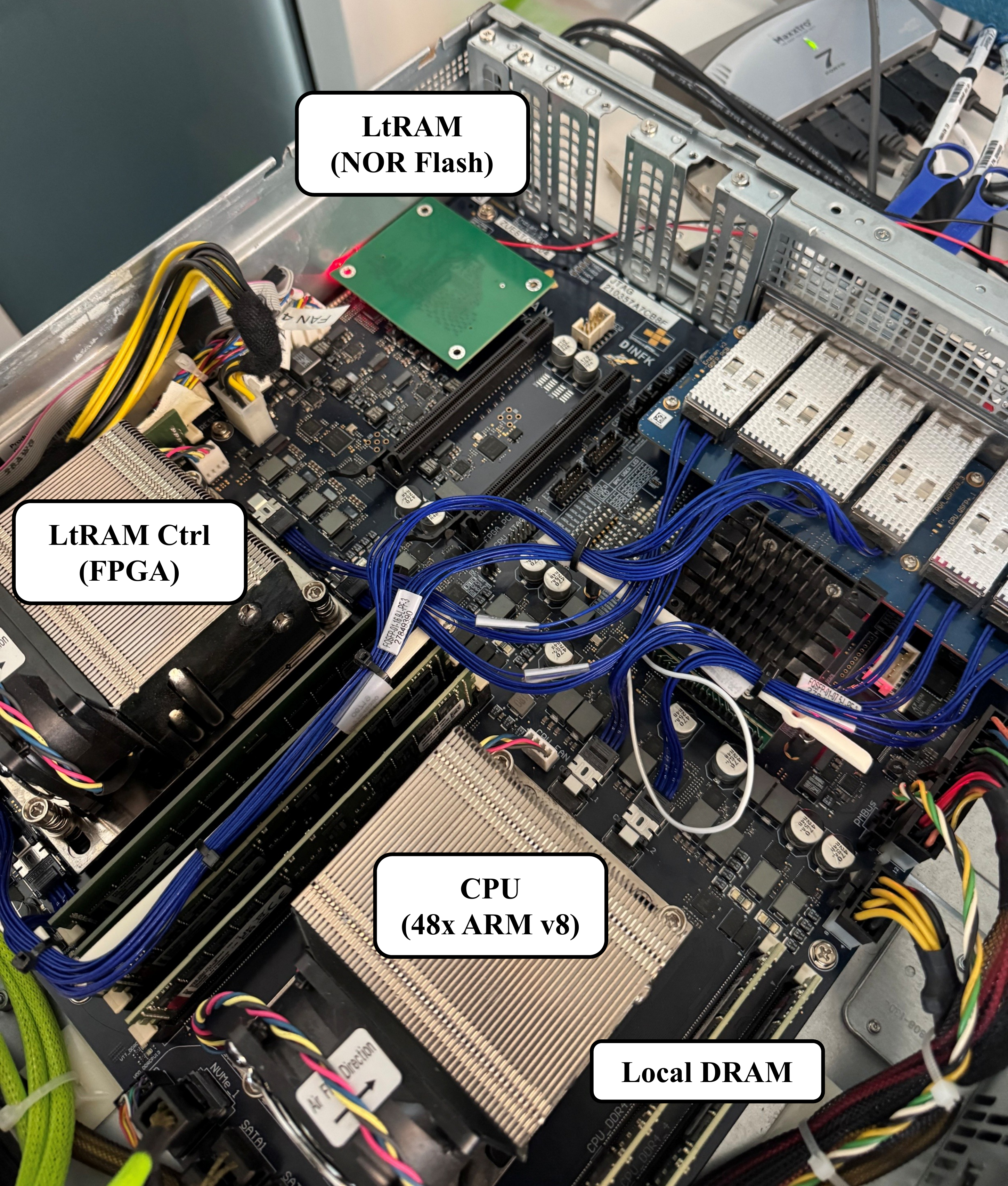}
  \caption{The Enzian research platform~\cite{enzian} with the NOR flash device attached. Photo from the Enzian project.}
  \label{fig:enzian}
\end{figure}

We prototype the interface on the Enzian research platform~\cite{enzian}: a server-class ARM CPU paired with a coherent FPGA fabric, configured here so the FPGA acts as the LtRAM memory controller and exposes LtRAM to the CPU through Enzian's cache-coherent interconnect. The CPU runs Linux kernel 6.8 with our page-migration extensions for LtRAM.

We chose NOR flash for the prototype for its availability and low cost: a 256\,MB Micron Serial NOR Flash device~\cite{micron_mt35xu02g} (Figure~\ref{fig:enzian}) drops onto the FPGA board, whereas an MRAM or RRAM prototype would have required a research-specific tape-out. NOR is not a canonical LtRAM technology, but it shares the asymmetry shape the interface targets: byte-addressable reads, 256\,B writes, 4\,KB erases, and $10^5$ erase cycles per block. Measured chip-level latencies are 490\,ns for a 128\,B read (Enzian's cache line), 16\,$\mu$s for a 256\,B write, and 18\,ms for a 4\,KB erase.
\subsection{Controller}
\label{sec:controller}


The FPGA-based memory controller implements the three responsibilities specified in \S\ref{sec:interface_controller}. The 4\,KB write granularity coincides with both the OS base page size and the NOR device's erase block size, eliminating controller-side merging.

Beyond exposing the interface, the controller adds two optimizations that preserve read performance under the read-mostly workloads LtRAM serves. First, incoming writes are absorbed into an SRAM buffer and acknowledged before the write is committed to NOR flash, so reads to other memory regions are not stalled by the slower device-level program. Second, the erase command is exposed to the OS rather than triggered internally, so the OS can schedule erases during periods of low read pressure. Everything else, including wear distribution and placement, lives in the OS.

Removing the AIT eliminates its on-DIMM DRAM cache, the media region reserved for AIT storage, and the lookup latency on the read critical path (\S\ref{sec:ait}). OS scheduling of erases and migrations further avoids the tail-latency from hardware wear-leveling (\S\ref{sec:wearleveling}). The controller retains only a small bad-block remap table whose coarse granularity (e.g., 64\,KB per entry in Optane) lets it reside entirely in SRAM.
\subsection{Operating System}
\label{sec:os}

The OS implementation realizes the four guarantees of \S\ref{sec:interface_os} through two mechanisms: memory-aware page placement 
and a token-based wear-leveling subsystem. They keep LtRAM writes within the device's endurance budget and scale to multi-terabyte capacities without per-page metadata bloat.

\subsubsection{Page Placement Policies}
The OS implements three placement policies: allocation, migration to LtRAM, and migration to DRAM.

\paragraph{Allocation}
LtRAM is implemented as a new Linux zone with its own free-page list; the allocator selects DRAM or LtRAM per request based on write intensity. Pages are placed directly on LtRAM only when the kernel or application can establish that the data is read-only by construction (executable code, shared libraries, read-only memory-mapped files). Everything else is first allocated in DRAM and may move to LtRAM later via the migration policy.

\paragraph{Migration to DRAM}
All LtRAM pages are write-protected.
On a write fault, the kernel performs a CoW migration: it allocates a DRAM page, copies the LtRAM page's contents, updates the page table, and executes the store.
The original LtRAM page is then freed and its underlying block scheduled for erase when the device is idle. 
All application-level writes execute on DRAM, preserving the AROM invariant.

\paragraph{Migration to LtRAM}
Only read-mostly pages should be migrated to LtRAM.
To determine which pages have not been written recently, we reuse existing page dirty bit tracking mechanisms in the kernel.
On every scan interval $T$, the kernel scans the target process’s PTEs: pages with a clear dirty bit become migration candidates, and the rest have their dirty bits cleared for the next scan. 
Candidates are migrated to LtRAM in scan order, subject to the endurance token budget (\S\ref{sec:token-allocator}). 
This approach is cheap but imprecise; better migration decisions are left to future work (\S\ref{sec:best_migration}).

\subsubsection{Wear-leveling by token allocation}
\label{sec:token-allocator}

Our token allocator releases tokens at a rate determined by the device's endurance budget and target deployment lifetime.
For an LtRAM with $N$ pages rated for $E$ erases each over a $T$-second deployment, the rate is $r = NE/T$ tokens per second.
Every migration to LtRAM must consume a token; the kernel defers a migration if no tokens are available. Tokens may be banked across idle windows, so an under-used workload can trade idle time for a later migration burst while the long-run average stays within budget by construction. Because the OS is the only writer and the token allocator paces those writes, the device meets its lifetime budget by construction without persisting per-block erase counts; only the running token balance is required.
This bounds total wear but does not enforce uniformity across the chip's physical sections; \S\ref{sec:distributing_wear} discusses how to achieve uniformity.

\subsubsection{What changes in the kernel?}
Our design reuses much of the page management machinery already present in the Linux kernel. Adding a new zone at the top of the zone fallback order separates the DRAM and LtRAM allocation paths and prevents allocations onto LtRAM unless explicitly requested, keeping write-heavy pages off LtRAM. The CoW fast path already handles fault-on-write semantics; we extend the trigger so LtRAM-resident pages are marked CoW by default and write faults migrate the contents to DRAM before modification. Migration to LtRAM requires the largest modifications, as it involves determining which pages should be migrated to LtRAM and implementing a token allocator.
\subsection{Projected Read Latency}
\label{sec:projections}

\begin{figure}[!t]
  \centering
  \includegraphics[width=\columnwidth]{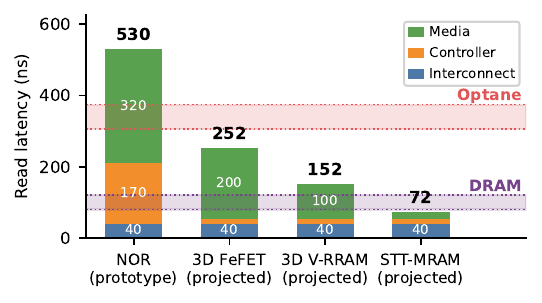}
  \caption{Projected read latency through proposed interface on a DDR-class interconnect, decomposed into interconnect, on-device controller, and media access. DRAM (80--120\,ns) and Optane (305--374\,ns) bands serve as reference points.}
  \label{fig:projected-latency}
\end{figure}

Figure~\ref{fig:projected-latency} projects end-to-end read latency for a 128\,B cache-line access through our HW\slash SW interface on a DDR-class interconnect, decomposed into interconnect, on-device controller, and media access. Media latencies come from Table~\ref{tab:ltram}. The interconnect is estimated at 40\,ns, the residual of DRAM's 80\,ns budget after subtracting a 30\,ns cell access~\cite{lee2009architecting} and a 10\,ns on-DIMM controller.

The on-device controller is based on our NOR prototype; we measure 490\,ns for controller and media combined. Media access accounts for 320\,ns, and the remaining 170\,ns is the controller, or 34 cycles at NOR's 200\,MHz clock. Reclocked to a DDR5-5600 clock (2.8\,GHz), the same 34-cycle budget drops to 12\,ns, which we adopt for the projected technologies. This estimate is conservative: 16 of the 34 cycles are wait states intrinsic to off-the-shelf NOR's command pipeline that a purpose-built LtRAM controller would not incur. The interconnect is held constant at 40\,ns across all technologies, assuming every candidate would sit on the same DDR bus.

All LtRAM technologies improve on Optane by 26--79\%, placing them closer to DRAM. Against DRAM's 80\,ns baseline, the NOR prototype projects to 6.6$\times$, 3D FeFET to 3.2$\times$, 3D V-RRAM to 1.9$\times$, and STT-MRAM to 0.9$\times$ (10\% faster than DRAM). STT-MRAM's projection is favorable; however, manufacturing and scalability challenges may prevent it from competing at DRAM capacity. The projections characterize single-access latency in isolation; at the application level, cache, prefetching, and memory-level parallelism overlap per-access costs, so workload-level performance on read-mostly data may approach DRAM more closely than the per-access numbers suggest. Application-level benchmark evaluation remains future work.

\section{Discussion}
\label{sec:discussion}

\subsection{Related Work}

\subsubsection*{DRAM-compatible NVM modules.} Intel Optane is the canonical example: a DRAM-compatible random-access interface backed by an on-DIMM translation layer~\cite{optane2022}. As \S\ref{sec:optane} showed, the translation layer inflates latency and module cost, eroding the cost advantage that made the memory device interesting. The shortcoming lies not in the underlying memory technology but in the interface.

\subsubsection*{Software-only tiering.}
\label{sec:tiering}
Google's software-defined far memory built on zswap~\cite{sdfm2019} shares the closest approach: OS manages a slower memory tier (compressed RAM) with the cost deferred to access time (decompression at fault).
HeMem~\cite{raybuck_hemem_2021}, TPP~\cite{tpp2023}, Memtis~\cite{lee_memtis_2023}, Colloid~\cite{colloid2024}, and Alto~\cite{alto2025} extend the same OS-managed-tier pattern to CXL-attached memory, but none reason about endurance, and all position the second tier as a slow tier between DRAM and storage.
Our design extends zswap's approach to a HW\slash SW co-design problem: LtRAM serves alongside DRAM as main memory rather than as a slow tier, and the deferred cost is a CoW migration paid only on writes to LtRAM pages rather than decompression on every access.

\subsubsection*{HW\slash SW co-design framings.} Microsoft's Managed Retention Memory work~\cite{mrm2025} argues that the right abstraction for emerging memory is neither storage class memory nor a DRAM replacement, but a category with its own interface.
Li et al.~\cite{li_towards_2025} laid out a similar split between long-term and short-term RAM.
Pond~\cite{pond2023} pushes in a related direction at the level of CXL pooling.
This paper builds on those framings and proposes a concrete HW\slash SW interface for LtRAM.

\subsection{Open Questions}

\subsubsection{What is the ideal LtRAM technology?}
\label{sec:ltram-tech}
Although the interface supports all LtRAM technologies, no candidate dominates on all axes today. Strong latency comes at the cost of density (MRAM), high density at the cost of endurance and write characteristics (RRAM, FeFET NAND), and manufacturing maturity at the cost of write granularity (PCM, as Optane showed). The harder problem is that most candidates have not yet scaled to production volume, so the tradeoffs themselves may shift as the memory technologies mature, and the answer will only become clearer as the memory devices scale.

\subsubsection{What is the best migration policy?}
\label{sec:best_migration}
The interface is not constrained to any single migration policy, but the right one remains open.
A dirty-bit-based classifier is the simplest and lowest-overhead option, but thresholding alone is too coarse: a page that passes the threshold in one observation window may be written in the next.
A small online predictor adapts to workload patterns from a few features (e.g., time since last write, recent write frequency) at low overhead, but introduces tuning overhead and needs feedback to learn which classifications were correct.
The right choice depends on the workload's diversity, the available per-page metadata budget, and how often workloads change at runtime.



\subsubsection{What is the best wear-leveling strategy?}
\label{sec:distributing_wear}
The token allocator distributes wear only across blocks in active use; blocks pinned by read-only pages stay outside the wear-leveling rotation.
Code binaries, ML weights, and similar load-time pages pin their blocks for the system's lifetime, causing the remaining blocks to wear out prematurely.
One solution is to evict pages from the least-worn blocks when the wear gap exceeds a threshold.
However, open questions remain: when to trigger eviction, what threshold to use, and how to pick candidates without inflating OS overhead.

%

\subsubsection{How can out-of-memory faults be prevented during write-heavy periods?}
\label{sec:dram_oom}
A workload phase change can flip a large LtRAM region from read-mostly to write-heavy all at once: every store triggers a CoW fault and a migration to DRAM, eventually exhausting the DRAM free list. Unlike CXL tiering~\cite{tpp2023,colloid2024,alto2025} where misplacement only slows reads, misplacement in LtRAM pays a per-store copy cost on the critical path. 
Admission control and DRAM-occupancy watermarks are starting points, but the right combination remains to be studied.

\section{Conclusion}
\label{sec:conclusion}

DRAM now exceeds half of server cost while its cost-per-bit has plateaued.
LtRAM is one solution, but DRAM-compatible interfaces incur too much overhead.
We propose a thin HW\slash SW interface that safely shifts controller responsibilities into the OS by enforcing a novel invariant we call Application Read-Only Memory (AROM): applications observe LtRAM as read-only memory, and only the kernel writes to LtRAM during page migrations.
The controller supports reads at cache-line granularity and writes at 4\,KB page granularity, and it exposes cell health metadata; the OS handles data placement, page migration, and wear-leveling.
We prototype this interface on the Enzian platform using NOR flash, with the goal of placing LtRAM alongside DRAM, serving as main memory for read-mostly data to lower DRAM footprint and per-server memory cost.

\bibliographystyle{ACM-Reference-Format}
\bibliography{dimes26}

\end{document}